\documentclass[journal]{IEEEtran}
\ifCLASSINFOpdf
\else
\fi
\usepackage{graphicx}
\usepackage{amsthm}
\usepackage{amsthm}
\usepackage{mathtools}
\usepackage{cite}
\usepackage{amsmath}
\usepackage{amssymb}
\usepackage{amsfonts}
\usepackage{amstext}
\usepackage{amsthm}
\usepackage{stfloats}
\usepackage{blindtext}
\usepackage{lipsum}
\usepackage[english]{babel}
\usepackage[utf8]{inputenc}
\usepackage{algorithm}
\usepackage{algpseudocode}
\usepackage{mathtools, nccmath}
\usepackage{color}
\usepackage{color,soul}
\input{insbox.tex}

\usepackage{changes}
\definechangesauthor[color=orange]{GS}
\definechangesauthor[color=blue]{KH}
\setremarkmarkup{(#2)}

\begin{document}
%
\title{Model Predictive Control Framework for Improving Vehicle Cornering Performance Using Handling Characteristics}
%
%
%

\author{Kyoungseok~Han~\IEEEmembership{Member,~IEEE}, Giseo~Park, Gokul S. Sankar, Kanghyun Nam,~\IEEEmembership{Member,~IEEE}, and Seibum B. Choi,~\IEEEmembership{Member,~IEEE}
\thanks{This work was supported in part by the National Research Foundation of Korea (NRF), funded by the Korean government (MSIP) under Grant 2017R1A2B4004116; in part by 
the BK21+ program through the NRF, funded by the Ministry of Education of Korea.}
\thanks{Kyoungseok Han, and Gokul S. Sankar are with University of Michigan, Ann Arbor, MI 48109 USA (email: kyoungsh@umich.edu, ggowrisa@umich.edu )

\indent Giseo Park, and Seibum B. Choi (Corresponding Author) are with KAIST, Daejeon, Korea (email: giseo123@kaist.ac.kr,  sbchoi@kaist.ac.kr)

\indent Kanghyun Nam (Co-corresponding Author) is with Yeungnam University, Gyeongsan, Gyeongbuk, Korea (Tel.: +82-53-810-2455; email: khnam@yu.ac.kr)}
}
\maketitle

\begin{abstract}
This paper proposes a new control strategy to improve vehicle cornering performance in a model predictive control framework. The most distinguishing feature of the proposed method is that the natural handling characteristics of the production vehicle is exploited to reduce the complexity of the conventional control methods. For safety's sake, most production vehicles are built to exhibit an understeer handling characteristics to some extent. By monitoring how much the vehicle is biased into the understeer state, the controller attempts to adjust this amount in a way that improves the vehicle cornering performance. With this particular strategy, an innovative controller can be designed without road friction information, which complicates the conventional control methods. In addition, unlike the conventional controllers, the reference yaw rate that is highly dependent on road friction need not be defined due to the proposed control structure. The optimal control problem is formulated in a model predictive control framework to handle the constraints efficiently, and simulations in various test scenarios illustrate the effectiveness of the proposed approach.
\end{abstract}


%
\IEEEpeerreviewmaketitle

\section{Introduction}
\IEEEPARstart{T}{he} demand for high-performance vehicles has increased recently \cite{hindiyeh2014controller}, and active vehicle chassis control systems have emerged, allowing for agile vehicle maneuvering \cite{guo2016nonlinear, de2014comparison}. For example, one representative technology called torque vectoring varies torque independently on each wheel to increase the vehicle dynamic performance. Since the vehicles with larger lateral acceleration for the same steering wheel angle have the ability to be dynamically driven \cite{de2018energy}, previous studies have attempted to increase lateral acceleration at a given steering angle. To achieve this objective, control methods based on yaw rate and/or sideslip have been suggested by many researchers \cite{di2013vehicle, lu2016enhancing, funke2017collision, nam2012lateral, hu2017robust}. The common objective of these two schemes is to maximize the lateral acceleration of the vehicle by making the most of the given road friction. In other words, when the vehicle reaches its handling limit, which is physically bounded by road friction, the control goal can be achieved. 
\subsection{Literature Review}
One method to increase the vehicle dynamic performance is to generate additional yaw moment by appropriately distributing the driving and/or braking torque to individual wheels. This is referred to as the ``yaw rate-based control" scheme, and the vehicle's yaw rate is controlled to track the reference yaw rate. Yaw rate-based control has been widely adopted due to its ease of implementation and intuitiveness. However, this method is not appropriate for the high-performance vehicle control. This method is usually recommended for vehicle stabilization when a certain amount of deviation of the current yaw rate from the reference value is detected \cite{nam2012lateral, hu2017robust}.

In addition to yaw rate-based control, other control schemes may be appropriate for maximizing the lateral acceleration, e.g., ``sideslip-based control''. Among these technologies, active front steering (AFS) directly modifies road wheel angles \cite{ren2016mpc,choi2016mpc}. Unlike in the yaw rate-based method, lateral tire force can be directly controlled to track the reference sideslip without intervention of longitudinal force. 
In general, AFS is combined with differential braking to enhance both cornering performance and vehicle stabilization. However, the very sophisticated coordinating of AFS and differential braking is still challenging and the extra cost of installing an AFS device is not practical for the automotive industry. Furthermore, estimating the sideslip required to implement this method is another significant challenge. 

As mentioned, the control performance is significantly affected by road friction. Since the maximum achievable vehicle lateral acceleration is physically bounded by road friction, accurate information on road surface conditions is the most useful information for designing the controller. That is, the more road surface information we have, the easier it is to control the vehicle's lateral acceleration to reach the friction limit. However, road friction information is not easily obtained in reality, and real-time estimation of road friction still also remains as an unresolved issue. Although there have been many efforts to construct the friction estimator, the robustness and accuracy of the estimators developed are not sufficient to be implemented in production vehicles \cite{han2017adaptive, hsu2010estimation}. \\ 

\subsection{Original Contributions}
As an approach to overcome the above-mentioned problems, we present an innovative control method that exploits the natural handling characteristics of production vehicles. More specifically, most production vehicles are designed to have understeer handling characteristics to prevent oversteer, which most critically needs to be avoided. The controller designed in this paper adjusts the amount of this understeer in such a way as to improve the cornering performance. To the authors' knowledge, such an approach based on this handling characteristics has not been previously reported for designing vehicle chassis controller. 

\begin{figure}[!t] 
\centering
\includegraphics[width=3.0in]{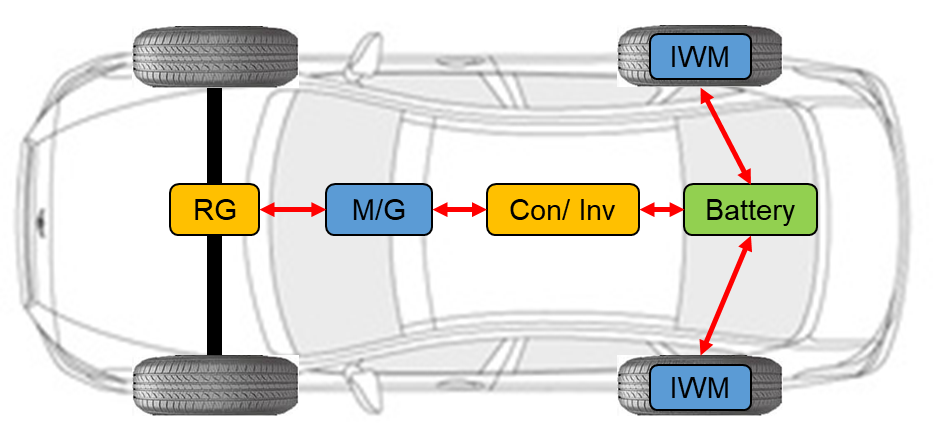} 
\caption{Powertrain configuration of the target vehicle (RG: reduction gear, M/G: motor/generator, and Con/Inv: converter/inverter.}
\label{vehicle_blueprints}
\end{figure}

As described in Fig. \ref{vehicle_blueprints}, the target vehicle is the pure electric vehicle with front wheel drive. However, in-wheel motors (IWMs) are additionally employed for the rear wheels, so an additional yaw moment can be applied to the vehicle. Therefore, the proposed method shares key strategies with yaw rate-based control, but it is distinguished in that the available road surface friction is utilized as much as possible by exploiting handling characteristics. Unlike a conventional yaw rate-based control, we do not explicitly define the desired yaw rate in this paper. Instead, the difference in wheel sideslip between the front and rear wheels is controlled to maintain a constant gap. As a result, the vehicle is steered to a nearly neutral steer condition in which the rear wheel sideslip increases up to a similar magnitude of the front wheel sideslip. Moreover, the proposed approach is applicable regardless of the type of road surface. Therefore, the most challenging task in controlling the vehicle, i.e., road friction estimation, can be completely excluded from the proposed approach.

The original contributions of this paper can be summarized as follows. (i) Using natural handling characteristics of production vehicles, an innovative control strategy that is robust to road friction is presented. (ii) Unlike a conventional yaw rate-based control, the proposed controller designed attempts to utilize road friction as much as possible to increase the vehicle dynamic performance. The controller is designed based on our newly established model that describes the wheel sideslip angle difference, which has not been reported so far in previous studies.
\subsection{Paper Layout}
The remainder of this paper is organized as follows. Section II briefly compares the proposed method with the conventional methods. We analyze vehicle handling characteristics in Section III. Section IV presents the control strategy, which is the central part of this paper. The effectiveness of the proposed method is verified by simulations in Section V, and we conclude the paper in Section VI.

\section{Comparison of Traditional and Proposed Control Methods}
This section discusses the inherent shortcomings of the traditional vehicle control methods, and briefly introduces the structure of the proposed controller to address the issues of the conventional methods. 

\begin{figure}[!t]
\centering
\includegraphics[width=3.0in]{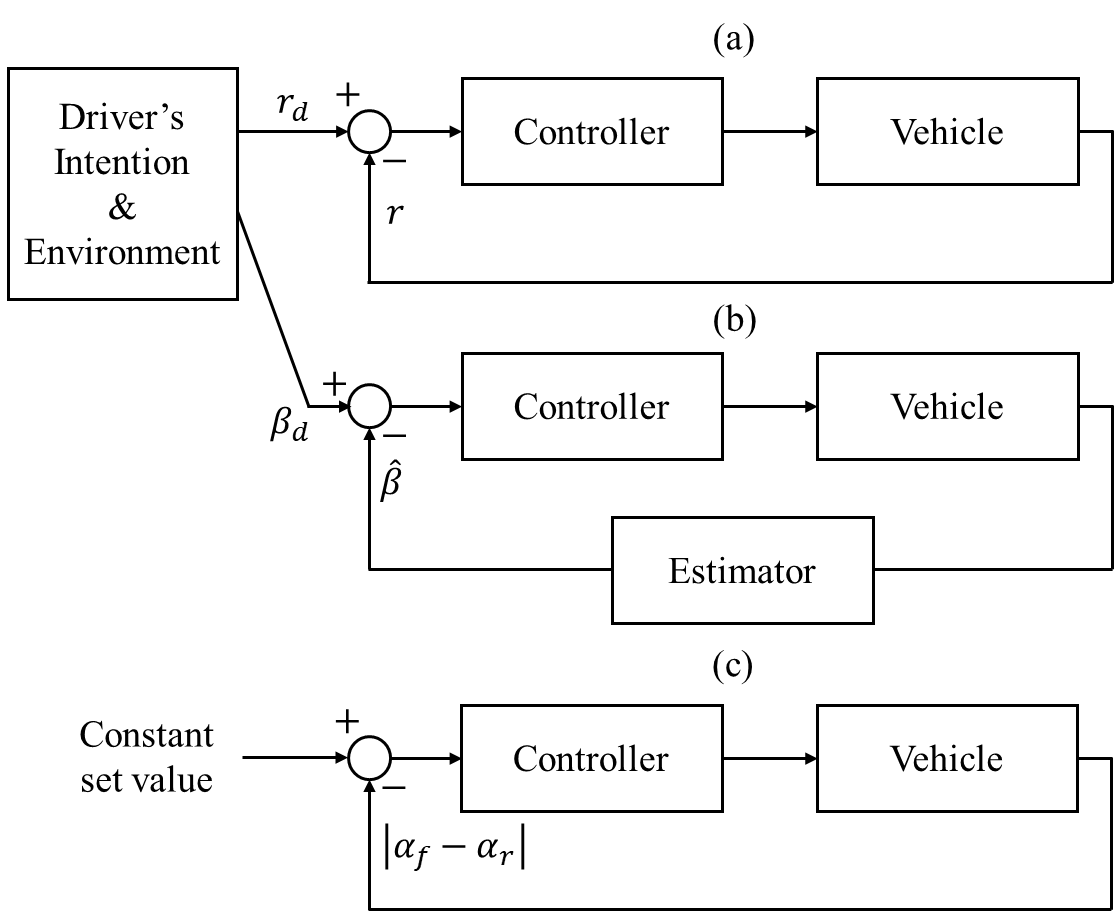}
\caption{Comparison of traditional and proposed control methods: (a) yaw rate-based control, (b) sideslip-based control, and (c) proposed method.}
\label{fig1}
\end{figure}

\textcolor{black}{Figure} \ref{fig1}(a) describes a block diagram of a conventional ``yaw rate-based control" approach. Since yaw rate measurement, $r$, is very accurate, with no offset, the controller can be easily designed by defining the desired yaw rate $r_{d}$. Depending on the driver's intention and the road environment, the desired yaw rate is often defined as follows \cite{rajamani2011vehicle}:
\begin{equation}\label{eq:eq1}
r_{d}=\frac{v_{x}}{L+ \underbrace {\left ( \frac{ml_{r}}{C_{f}L}-\frac{ml_{f}}{C_{r}L} \right )}_{\kappa}{v_{x}}^{2}}\leq \frac{\mu g}{v_{x}}.\\
\end{equation}
where $v_{x}$ is the vehicle speed, $m$ is the vehicle \textcolor{black}{mass}, $l_{f}$ and $l_{r}$ are the distances from the center of gravity to the front and rear axles, $C_{f}$ and $C_{r}$ are the front and rear tire cornering stiffness values, $\mu$ is the road friction coefficient, $g$ is the gravitational constant, and $L=l_{f}+l_{r}$.

Traditionally, the parameter $\kappa$, called the understeering gain \cite{gillespie1992fundamentals}, determines the dynamic characteristics of the vehicle. If the constant $\kappa$ is specified using the nominal values of tire cornering stiffness, and the controller is designed to track the $r_d$, a steady-state cornering condition can be achieved. Therefore, the lateral instability of the vehicle can be stabilized by this nominal $\kappa$. In contrast, $\kappa$ can be updated online to increase the vehicle cornering performance. In theory, the lateral acceleration of the vehicle can be increased up to the road friction limit, as described in \eqref{eq:eq1}. Therefore, if $\mu$ is given accurately in real-time, the desired yaw rate that maximizes the vehicle lateral acceleration with the given driver's inputs can be easily defined:
\begin{equation}
    r_d^{\text{max}}=0.85\cdot\frac{\mu g}{v_{x}}.
\end{equation}
Here, 0.85 is set as the safety margin \cite{rajamani2011vehicle}, and the controller can be designed to minimize the error between $r$ and $r_d^{\text{max}}$. Because $r$ is a precisely measurable value, this is very straightforward control structure. However, estimation of $\mu$ is very difficult and still a challenge in the field of vehicle dynamics research. For these reasons, it is very important to determine the appropriate desired yaw rate, which depends heavily on the driver's inputs and road friction. 

An even better solution to maximizing the utilization of surface friction is to design a ``sideslip-based controller'', as described in Fig. \ref{fig1}(b). Since sideslip $\beta$ is directly related to the tire lateral force, the appropriate control of $\beta$ is a promising method for enhancing cornering performance. However, sideslip-based control scheme in Fig. \ref{fig1}(b) requires the development of a $\beta$ estimator, which requires huge efforts \cite{nam2013estimation, park2018integrated}. Further exacerbating the situation, the desired sideslip $\beta_{d}$ cannot be easily determined due to the inaccuracy of the road friction information, which is the case in yaw rate-based control. Therefore, when designing $\beta$-based control, neither the desired state variable nor the current state variable is given. For these reasons, the vehicle controller is traditionally designed by heuristically tuning with the yaw rate-based and sideslip-based control together.

To overcome the above mentioned issues, we present a new method that does not need to determine the desired state variable. As exhibited in Fig. \ref{fig1}(c), the desired constant value is always specified as a constant value, and the feedback value is the wheel sideslip difference between the front and rear wheels, i.e., $\alpha_{f}-\alpha_{r}$, which can be measured in real-time \cite{han2018estimation}. The set value is specified as constant regardless of road frictions, and hence the proposed method is effective on any type of road surface. This is possible because we exploit the natural handling characteristics of the production vehicle; more details are introduced in the following sections.

In this paper, the problem is formulated in the framework of model predictive control (MPC) \cite{borrelli2017predictive}, which can solve the constrained optimal control problem. In general, the nonlinear controller is effective for systems with constraints. MPC is a general, systematic and flexible nonlinear controller for constrained systems based on prediction and optimization. Basically, MPC uses the receding horizon control principle, which has the ability to predict a future response and can accordingly take the best control action at the current time slot. Based on a model of the system dynamics, MPC computes the optimal input profile during a finite time horizon with respect to a specified performance index. This process for calculating the control input profile is repeated every control cycle when new information on the system is updated. Due to this property, MPC needs an excessive amount of computational effort compared to the classical control approach. Therefore, over the past few years, MPC has only been applied to systems with slow dynamics, such as in process control industries. However, the recent development of computational algorithms and numerical analysis techniques enables MPC to be applied to systems with fast dynamics such as the vehicle dynamic control research area \cite{beal2013model, del2010automotive, borrelli2006mpc}, and we also employ MPC to ensure the vehicle stability by enforcing the state and control within acceptable bounds.
\section{Vehicle Cornering Dynamics and Handling Characteristics}
\subsection{Vehicle Cornering Dynamics}
\begin{figure}[!t]
\centering
\includegraphics[width=3.0in]{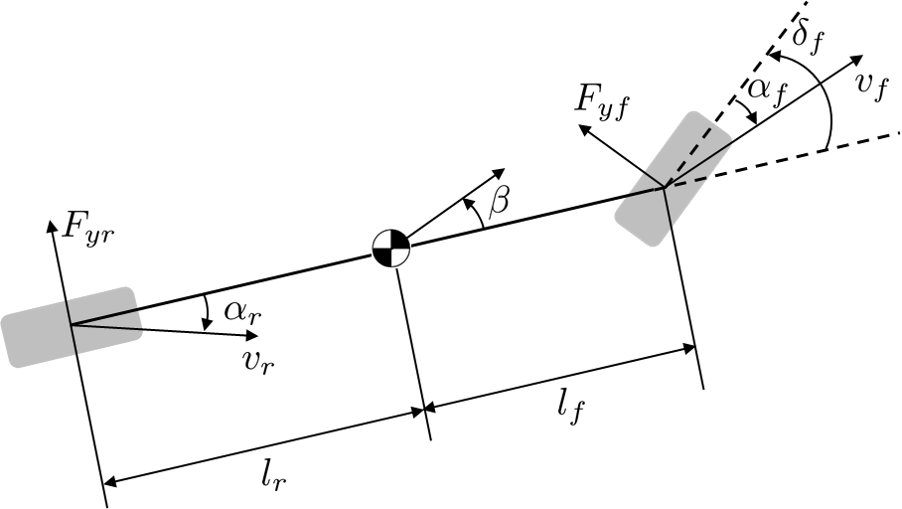}
\caption{\textcolor{black}{Bicycle model.}}
\label{fig2}
\end{figure}

The vehicle cornering dynamics can be approximated by the bicycle model in Fig. \ref{fig2}, which captures key cornering dynamic characteristics. By assuming a constant vehicle speed, which is a reasonable assumption in normal vehicle turns, the bicycle model is defined with state vector $x=[\beta, r]'$as follows \cite{rajamani2011vehicle}:
\begin{equation}\label{eq:bicyle_model1}
\dot{\beta}=\frac{F_{yf}\textcolor{black}{\text{cos}\delta_f}+F_{yr}}{mv_{x}}-r,
\end{equation}
\begin{equation}\label{eq:bicyle_model2}
\dot{r}=\frac{l_{f}F_{yf}\textcolor{black}{\text{cos}\delta_f}-l_{r}F_{yr}+M_{z}}{I_{z}}.
\end{equation}
where $\beta$ is the body sideslip, $F_{yf}$ and $F_{yr}$ are the front and rear axle lateral forces, $I_{z}$ is the yaw moment of inertia, $M_{z}$ is the corrective yaw moment, \textcolor{black}{$\delta_f$ is steered wheel angle, and we assume \text{cos}$\delta_f \approx 1$ using small angle approximation}.

For small tire slip angles, the lateral tire forces are expressed by a piecewise affine function \cite{di2013vehicle}:
\begin{equation}\label{eq:Fyf}
F_{yj}(\alpha_j)=
\begin{cases}
\begin{aligned}
& d_j(\alpha_j+\alpha_{j}^p)-e_j,& &\text{if  } \alpha_j<-\alpha_{j}^p  \\
& \textcolor{black}{-}C_{j}\alpha_{j},& &\text{if  } -\alpha_{j}^p \leq \alpha_j \leq \alpha_{j}^p\\
& d_j(\alpha_j-\alpha_{j}^p)+e_j,& &\text{if  } \alpha_j>\alpha_{j}^p
\end{aligned}
\end{cases}
\end{equation}
where $j\in\left \{ f, r \right \}$, $f$ and $r$ represent the front and rear tires, $\alpha_{j}$ is the wheel sideslip angle, and $C_j$ is the tire cornering stiffness, which is highly dependent on road friction and normal tire force. $\alpha_{j}^p$ is the point at which the lateral force begins to saturate and \textcolor{black}{wheel sideslip angle increases rapidly, which eventually leads to vehicle instability}, $d_j$ and $e_j$ are determined experimentally. 

Since the purpose of this paper is to improve vehicle cornering performance within the stable region, we mainly exploit a linear tire model, i.e., $F_{yj}=C_j\alpha_j \: (|\alpha_j|\leq\alpha_{j}^p)$, in the following sections. When the $\alpha_j$ is about to go above the saturation point $\alpha_{j}^p$, the vehicle stability control system should be activated to reduce the excessive $\alpha_{j}$. However, this stability issue is out of the scope of this paper and the designed controller here attempts to prevent this stability problem in advance by adopting the predictive control. Specifically, the state and control constraints are enforced to avoid the excessive wheel sideslip in advance. 

Because the vehicle is eventually controlled by the combination of applied tire forces, the state variable $\alpha_j$ can be used to more intuitively express the objective of the proposed approach. Therefore, the state variables defined in \eqref{eq:bicyle_model1} and \eqref{eq:bicyle_model2} are represented in terms of the wheel sideslip angles. 

In high-speed turns, the wheel sideslip angles in the stable region are approximated by \cite{rajamani2011vehicle}:
\begin{subequations}\label{eq:alpha_f}
\begin{align}
\alpha_{f}&=\beta+\frac{l_{f}}{v_{x}}r-\delta_{f},\\
\alpha_{r}&=\beta-\frac{l_{r}}{v_{x}}r.
\end{align}
\end{subequations}

From \eqref{eq:alpha_f}, the wheel sideslip angle difference between the front and rear wheels is easily calculated as follows:
\begin{equation}\label{eq:alpha_diff}
\alpha_{f}-\alpha_{r}=\frac{l_{f}+l_{r}}{v_{x}}r-\delta_{f}.
\end{equation}
\indent Note that \eqref{eq:alpha_diff} does not depend on the body sideslip $\beta$, which cannot be measured in production vehicles. Therefore, it is enough to express \eqref{eq:alpha_diff} with the standard sensor signals in production vehicles and constant parameter values. This is the basis of our approach. Moreover, the vehicle handling characteristics that is an essential part of the proposed approach is also determined by the sign of wheel sideslip difference, i.e., sign($|\alpha_f|-|\alpha_r|$).

Differentiating \eqref{eq:alpha_f} with respect to time leads to:
\begin{subequations}
\begin{align}
\label{eq:alpha_f_dot}
\dot{\alpha}_{f}&=\dot{\beta}+\frac{l_{f}}{v_{x}}\dot{r}-\dot{\delta}_{f},\\
\dot{\alpha}_{r}&=\dot{\beta}-\frac{l_{r}}{v_{x}}\dot{r}.
\end{align}
\end{subequations}

Substituting \eqref{eq:bicyle_model1}, \eqref{eq:bicyle_model2}, \eqref{eq:Fyf} and \eqref{eq:alpha_diff} into the above equations gives the following dynamic models in terms of wheel sideslip angles:
\begin{align}\label{eq:alpha_f_dot_model}
\dot{\alpha}_{f}=& &\left ( -\frac{C_{f}}{mv_{x}}-\frac{l_{f}^{2}C_{f}}{I_{z}v_{x}}-
\frac{v_{x}}{L} \right )\alpha_{f}+\left ( -\frac{C_{r}}{mv_{x}}+\cdots \right .\\
& &\cdots \left . \frac{l_{f}l_{r}C_{r}}{I_{z}v_{x}}+\frac{v_{x}}{L} \right )\alpha_{r}+\frac{l_{f}}{I_{z}v_{x}}M_{z}-\frac{v_{x}}{L}\delta_{f}-\dot{\delta}_{f} \\
\end{align}
\begin{align}\label{eq:alpha_r_dot_model}
\dot{\alpha}_{r}=&\left ( -\frac{C_{f}}{mv_{x}} \textcolor{red}{+}\frac{l_rl_fC_{f}}{I_{z}v_{x}}-
\frac{v_{x}}{L} \right )\alpha_{f}+\left ( -\frac{C_{r}}{mv_{x}}\textcolor{red}{-}\cdots \right .\\
&\cdots \left . \frac{l_{r}^2C_{r}}{I_{z}v_{x}}+\frac{v_{x}}{L} \right )\alpha_{r}-\frac{l_{r}}{I_{z}v_{x}}M_{z}-\frac{v_{x}}{L}\delta_{f}
\end{align}
Here, the control input is $M_z$, and $\delta_f$ is purely determined by the driver's input. Furthermore, $v_x$ is assumed to be constant, and the cornering stiffness $C_{f,r}$ can be estimated by our previous study \cite{han2018estimation} and is assumed to be constant. Therefore, in this model, the only varying parameter is the steering wheel angle.

In the next sub-section, the vehicle handling characteristics are categorized based on these equations of the motion.
\subsection{Vehicle Handling Characteristics}
The vehicle handling characteristics are classified into three types at steady-state cornering: understeer, oversteer, and neutral steer conditions \cite{rajamani2011vehicle}. In general, these handling characteristics are distinguished by the difference between the front and rear wheel sideslip angle as follows.
\begin{figure}[!t]
\centering
\includegraphics[width=3.0in]{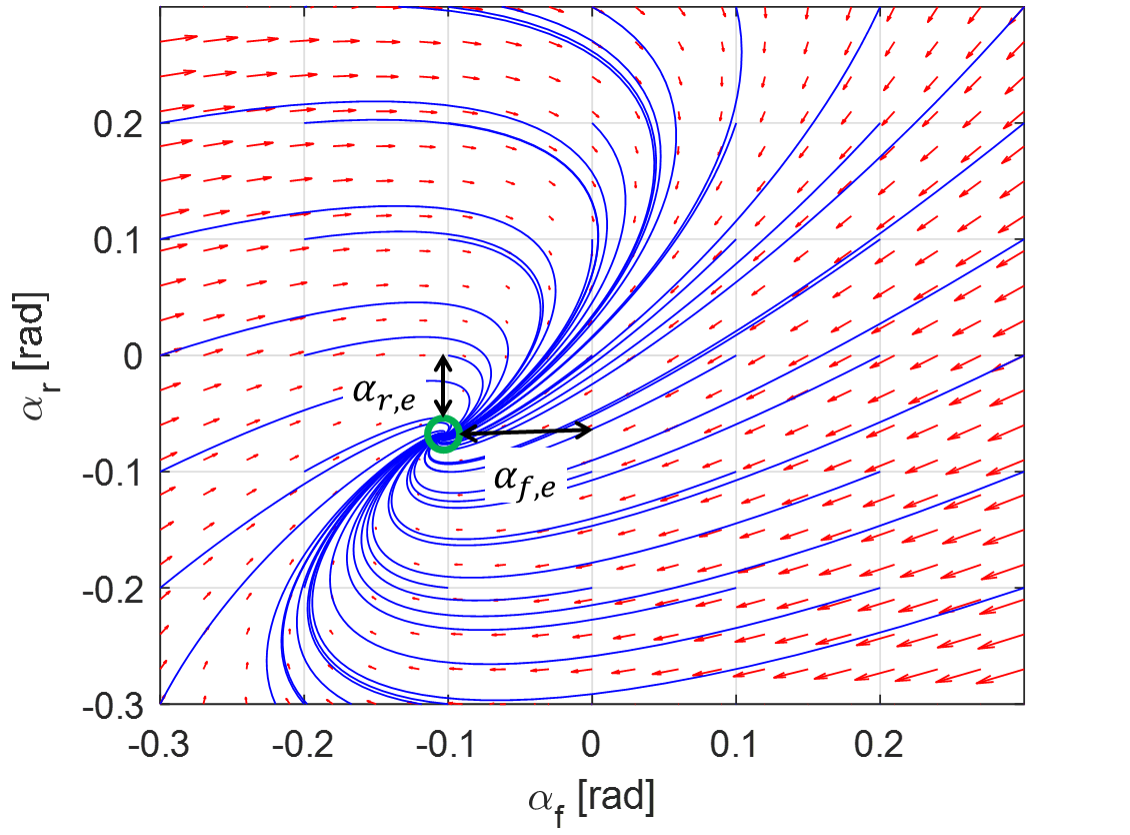}
\caption{Phase plot of the front and rear wheel sideslip angles.}
\label{phase_plane}
\end{figure}

For safety reasons, most production vehicles are built to exhibit some understeer characteristics, i.e., $\left |\alpha_{f}  \right |$$>$$\left |\alpha_{r}  \right |$. That is, when the vehicle is steered, the front wheel sideslip angle increases faster than the rear one. This understeer-biased handling is allowed to a certain extent during actual cornering.

In contrast, the oversteer condition, $\left |\alpha_{f}  \right |$$<$$\left |\alpha_{r}  \right |$, which should be avoided in most real driving, causes severe lateral instability of the vehicle if this condition persists. Therefore, the vehicle stability control system should be activated at this moment to reduce the excessive rear wheel sideslip.

And finally, neutral steer, $\left |\alpha_{f}  \right |$$=$$\left |\alpha_{r}  \right |$, is the ideal handling condition to improve cornering performance, but it is very difficult to achieve this condition all the time. Therefore, most production vehicles have been designed to show understeer handling characteristics.

Fig. \ref{phase_plane} provides the phase plot of the front and rear wheel slip angles based on \eqref{eq:alpha_f_dot_model} and \eqref{eq:alpha_r_dot_model} with the constant $\delta_f \approx 2.09 \text{ [rad]}$. The control input, namely, the additional yaw moment, is not applied to see the natural vehicle handling characteristics. As shown in Fig. \ref{phase_plane}, we can verify that the equilibrium point \cite{khalil2002nonlinear} is located where the absolute value of the front slip angle is larger than the rear one:
\begin{equation}\label{equilibrium}
|\alpha_{f,e}|>|\alpha_{r,e}|.
\end{equation}
where subscript $e$ denotes the equilibrium.

Although the vehicle occasionally exhibits oversteer or neutral steer conditions during transient maneuvers, the wheel sideslip angles eventually converge to the understeer area. This paper aims to modify this understeer-biased characteristics by applying additional yaw moment, and hence the vehicle can be controlled to display a nearly neutral steering.
\section{Control Design}
\subsection{Objective}
\begin{figure}[!t]
\centering
\includegraphics[width=3.5in]{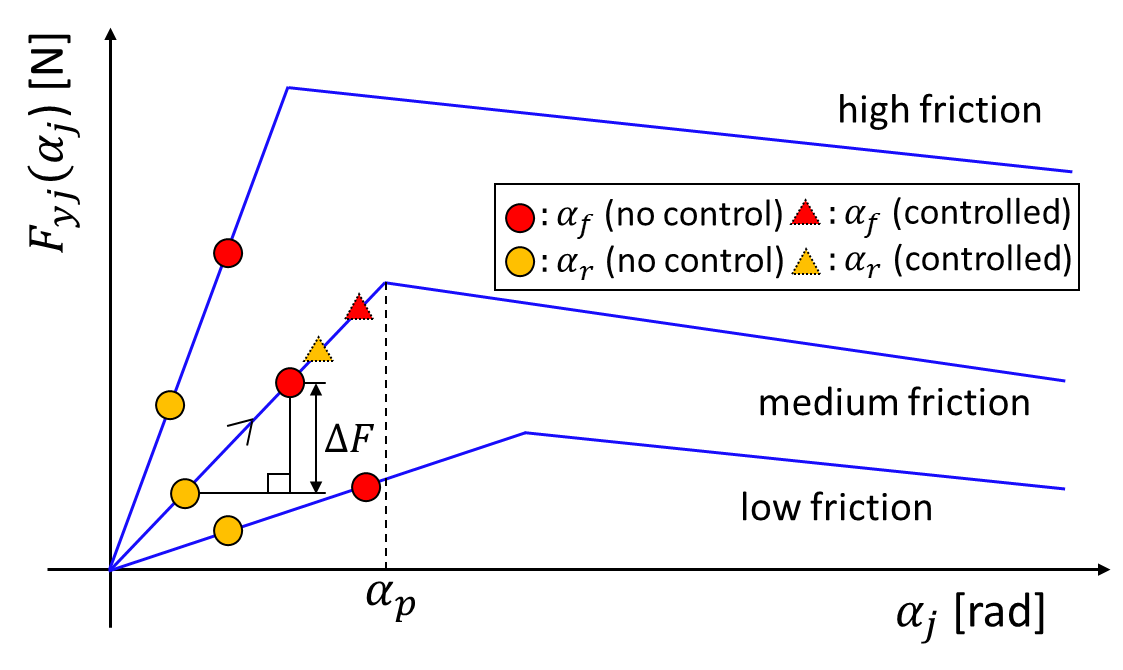}
\caption{Behavior of front and rear sideslip angles during normal turning on various road surface conditions.}
\label{alpha_diff}
\end{figure}

Figure \ref{alpha_diff} illustrates the behavior of the front and rear wheel sideslip angles during normal turning on various road frictions. As mentioned, the vehicle exhibits understeer-biased handling characteristics for all cases, i.e., $\left | \alpha_{f} \right |>\left | \alpha_{r} \right |$, regardless of the type of road surface. For example, in the case of a medium road friction surface, there is room for increasing the rear lateral tire force, i.e., $\Delta F$, while maintaining the vehicle stability. Given driver input $v_{x}$ and $\delta_{f}$, the sum of lateral tire forces can be maximized when the vehicle is steered to a neutral handling condition, i.e., $\alpha_{f}=\alpha_{r}$. Therefore, we can draw this potential force up to its limit. More specifically, the wheel slip angle difference between the front and rear wheels is controlled to maintain a constant gap because the exact neutral condition can easily cause vehicle instability by the unexpected uncertainties, as shown in Fig. \ref{alpha_diff}. In fact, both sideslip angles increase together due to the applied yaw moment, as depicted in Fig. \ref{alpha_diff}. Therefore, the control objective can be defined as follows:
\begin{equation}\label{eq:goal}
\textcolor{black}{\text{sign}\left ( \alpha_{f} -\alpha_{r} \right )\left ( \alpha_{f} -\alpha_{r} \right )\rightarrow \zeta.}
\end{equation}
\textcolor{black}{where $\zeta$ is a positive constant safety margin.}
\begin{figure*}[b]
\normalsize
\hrulefill
\begin{equation}
\label{matrices}
\begin{aligned}
\\
  &A_c= \begin{bmatrix}
-\frac{l_fC_f}{I_zv_x}L & \frac{L}{I_zv_x}\left ( -l_fC_f+l_rC_r \right ) \\ 
-\frac{C_f}{mv_x}+\frac{l_fl_rC_f}{I_zv_x}-\frac{v_x}{L} & -\frac{C_f+C_r}{mv_x}+\frac{l_fl_rC_f}{I_zv_x}-\frac{l_r^2C_r}{I_zv_x}
\end{bmatrix},\quad 
B_c=\begin{bmatrix}
\frac{L}{I_zv_x}\\ -\frac{l_r}{I_zv_x}
\end{bmatrix},\\ \\
&E_c=\begin{bmatrix}
-\dot{\delta}_f\\ -\frac{v_x}{L}\delta_f
\end{bmatrix}, \quad
C_c=\begin{bmatrix}
\frac{v_x}{L} & 0 \\  
-\frac{C_f}{m} & -\frac{C_f+C_r}{m}
\end{bmatrix}, \quad
D_c=\begin{bmatrix}
\frac{v_x}{L}\delta_f\\ 0
\end{bmatrix}.
\end{aligned}
\end{equation}

\vspace*{4pt}
\end{figure*}

If $\alpha_f$ is about to go into an unstable region ($\alpha_f > \alpha_{f}^p$) or change to an oversteering condition ($\alpha_f-\alpha_r<0$), the stability control system must be activated. We can monitor the vehicle stability status by calculating the magnitude of $\alpha_f-\alpha_r$ or estimating $\alpha_f$ in real-time, which provides the activation timing of the stability system. In order to ensure the safety, the current $\alpha_f$ should be less than an allowable wheel sideslip angle bound $\alpha_{f}^{\text{max}}$ regardless of road type. This constraint is specified in our optimal control problem. In short, the defined models in \eqref{eq:alpha_f_dot_model} and \eqref{eq:alpha_r_dot_model} are enforced to the region where $\alpha_f< \alpha_{f}^{\text{max}}$ and $|\alpha_f|>|\alpha_r|$, and the other regions are not in the scope of this paper.

\subsection{Control Oriented Model}
In this sub-section, the developed cornering dynamics model in terms of wheel sideslip angles in \eqref{eq:alpha_f_dot_model} and \eqref{eq:alpha_r_dot_model}  are manipulated to explicitly express the purpose of this paper. Subtracting \eqref{eq:alpha_r_dot_model} from \eqref{eq:alpha_f_dot_model} and manipulating the equations gives the following new system dynamics:
\begin{equation}\label{state_space}
\dot{x}(t)=A_cx(t)+B_cu(t)+E_c(t),
\end{equation}
\begin{equation}\label{measure}
y(t)=C_cx(t)+D_c(t),
\end{equation}
where $x(t)\in\mathbb{R}^2$, $y(t)\in\mathbb{R}^2$, $u(t)\in\mathbb{R}$ are the state, output and control input vectors, respectively.

The state vector in \eqref{state_space} is  $x=[x_1,\: x_2]'=[\alpha_f-\alpha_r,\: \alpha_r]'$, the output in \eqref{measure} is $y=[r,\: a_y]'$, and the control input is the additional yaw moment $M_z$ as follows:
\begin{equation}\label{Mz}
    M_z = \frac{l_w}{2 r_w}(T_{m,r}-T_{m,l}).
\end{equation}
where $r_w$ is the tire radius, $l_w$ is the track width, $T_{m,r}$ and $T_{m,l}$ are the applied motor torque at the rear right and rear left wheels, respectively. The matrices $A_c,\: B_c \text{ and } C_c$ are time invariant, but $E_c \text{ and } D_c$ vary depending on driver's input $\delta_f$, as defined in \eqref{matrices}.

It should be noted that the defined models here were first introduced in related studies, and we can expect the behaviors of sideslip difference $x_1$ and rear wheel sideslip $\alpha_r$ according to control input $u$ based on these models.

As mentioned, the state variable $x_1$ is readily available by \eqref{eq:alpha_diff}. However, it is difficult to measure the state $x_2$ directly, so an estimation method should be considered to fully exploit the developed model. 

In fact, numerous efforts have been made to estimate the vehicle sideslip in the literature \cite{oh2012vehicle}. Since $x_2$ is a function of $\beta$ as described in \eqref{eq:alpha_f}, we address $\beta$ estimation instead of $\alpha_r$, which is a more general expression. In the literature, the main effort has been made to robustly estimate $\beta$ even if the vehicle is traveling at the unstable region, e.g., $|\alpha_f|<|\alpha_r|$ or \textcolor{black}{$|\alpha_{f}^p|<|\alpha_f|$}. The reason $\beta$ estimation is challenging is that the matrices $A_c,\: B_c \text{ and } C_c$ are still not constant values when the tire slip angle begin to be saturated. At that time, the tire cornering stiffness $C_{f,r}$ reduces nonlinearly. Therefore, additional cornering stiffness adaptation is often considered  \cite{you2009new}. However, this compensation may not always capture the accurate variation of cornering stiffness, so this additional task generally reduces model accuracy in \eqref{state_space} and \eqref{measure}. 

However, as mentioned, our interest area of sideslip angles in the tire force curve is the stable region, i.e., $-\alpha_{j}^p \leq \alpha_j \leq \alpha_{j}^p$, so the assumption that $C_{f,r}$ is constant can accompany it. Therefore, a typical Luenberger observer can be developed to estimate $x_2$ as follows \cite{oh2012vehicle}:
\begin{equation}\label{observer}\setcounter{equation}{20}
\dot{\hat{x}}(t)=A_c\hat{x}(t)+B_cu(t)+E_c(t)+l \cdot \underset{y-\hat{y}}{\underbrace{C_c(x-\hat{x})}} .
\end{equation}
where $l\in\mathbb{R}^{2\times2}$ is the observer gain matrix, and $\hat{x}=[\hat{x_1}, \hat{x_2}]'$ is the estimated state variable.\\
\indent The system stability can be guaranteed by choosing the appropriate observer gain matrix. The interested readers can refer to \cite{oh2012vehicle, oh2011design}. Now the system in \eqref{state_space} is full state available system, and the optimal control problem can be formulated in MPC framework.
\subsection{MPC Problem Formulation}
\begin{figure}[!t]
\centering
\includegraphics[width=3.0in]{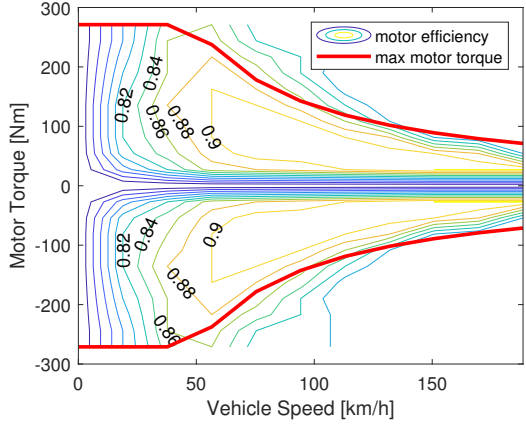}
\caption{Motor efficiency and maximum and minimum motor torque according to vehicle speed.}
\label{motor_map}
\end{figure}
In this sub-section, the optimal control problem to be solved is formulated in a MPC framework. As mentioned, the control objective of the proposed approach is to keep the desired constant gap for the wheel sideslip difference while maintaining the rear wheel sideslip within acceptable bounds.\\
\indent To ensure the control input within physically admissible ranges for all actuation times, the following motor torque range is specified:
\begin{equation}\label{control_constraint}
     \frac{T_{m}^{\text{min}}(v_x)}{r_w} l_w \leq u(t)\leq \frac{T_{m}^{\text{max}}(v_x)}{r_w} l_w.
\end{equation}
where $u(t)$ is the corrective yaw moment, and $T_{m}^{\text{max}}(v_x)$ and $T_{m}^{\text{min}}(v_x)$ are the maximum and minimum motor torques according to the vehicle speed, as depicted in Fig. \ref{motor_map}.
\indent The state variables also needs to be constrained by the following inequality condition to avoid \textcolor{black}{the tire force saturation:}
\begin{equation} \label{state_constraint}
\begin{bmatrix}
\zeta^{\text{min}} \\ \alpha_r^{\text{min}} 
\end{bmatrix}
\leq 
\begin{bmatrix}
x_1(t)\\x_2(t) 
\end{bmatrix}
\leq
\begin{bmatrix}
\zeta^{\text{max}} \\ \alpha_r^{\text{max}} 
\end{bmatrix}.
\end{equation}
where $x_1(t)=\alpha_f(t)-\alpha_r(t)$, and $x_2(t)=\alpha_r(t)$.\\
\indent The limits $\zeta^{\text{min}}$ and $\zeta^{\text{max}}$ are the maximum and minimum allowable distance between the front and rear wheel sideslip angles, and the rear wheel sideslip angle is additionally constrained by $\alpha_r^{\text{min}}$ and $\alpha_r^{\text{max}}$ to keep the vehicle within the stable region.

To formulate the MPC framework, the defined models in \eqref{state_space} and \eqref{measure} are discretized using the Euler's method with sampling period $T_s$ as:
\begin{subequations} \label{developed_model}
\begin{align} 
    x(k+1)&=A_dx(k)+B_du(k)+E_d,\\
    y(k)&=C_dx(k)+D_d.
\end{align}
\end{subequations}
\indent Using these discrete models, the MPC problem can be formulated as follows: 

\begin{align}\nonumber
     &\textcolor{black}{\min_{U_N(t)} J=\sum_{k=0}^{N-1}  \left \| x(k+1|t)-x_r(t) \right \|_Q^2+\left \| u(k|t) \right \|^2_R} \label{MPC_problem} \\
     & \qquad \qquad \qquad \textcolor{black}{+ \left \| \Delta u(k|t) \right \|^2_W} \\
     &\quad \quad \textrm{s.t.} \nonumber
\end{align}
\begin{subequations}
    \begin{align}
    x(k+1|t)&=A_dx(k|t)+B_du(k|t)+E_d\\
    \textcolor{black}{x(0|t)}&\textcolor{black}{=x(t)} \label{x_meas}\\
    &u^{\text{min}} \leq u(k|t) \leq u^{\text{max}}\\
    &\textcolor{black}{x^{\text{min}} \leq x(k+1|t) \leq x^{\text{max}}} \label{x_const}\\
    &\qquad \qquad k=0,...,N-1, \nonumber
    \end{align}
\end{subequations}
where \textcolor{black}{expression $\left \| x\right\|^2_Q$ means $x^TQx$},  $N$ is the time horizon, $\Delta u(k|t)=u(k|t)-u(k-1|t)$ is the control change rate included to prevent a large control change. \textcolor{black}{Here, the control change rate is described as: $\Delta u(0|t)=u(0|t)-u(-1|t)$ at initial time step $k=0$, and $u(-1|t)$ illustrates the previous control one step before, i.e., $u(t-1)$. $x(t)$ in \eqref{x_meas} is the estimated and measured state vector at step $t$ that gives the feedback functionality. }The predicted state variables over the time horizon, i.e., $x_1(k+1|t) \text{ and } \hat{x}_2(k+1|t)$, are constrained between $x^{\text{min}} \text{ and } x^{\text{max}}$ in \eqref{x_const}, $x_r(t)=[\zeta \quad \alpha_r^{\text{des}}]'$ is the constant desired values for states, but $\alpha_r^{\text{des}}$ is not explicitly assigned because no road friction information is given. Instead, $\alpha_r^{\text{des}}$ is specified as an appropriate constant value according to its sign to assign the correct direction of each motor torque in the initial phase.

In addition, $ Q\in\mathbb{R}^{2\times2} , R\in\mathbb{R}, \text{and }W\in\mathbb{R} $ are non-negative weighting matrices. \textcolor{black}{The weight matrix for the state vector is specified as, $Q=\text{diag}(Q_{11}, Q_{22})$. Since the main purpose of the controller is to track the desired value of $x_1(k+1|t)$, weighting $Q_{11}$ should be set to be larger than $Q_{22}$, i.e., $Q_{11}>>Q_{22}$. This is reasonable because $x_1(k+1|t)$ and $x_2(k+1|t)$ are highly related to each other as \eqref{state_space}, so if $x_1(k+1|t)$ is properly controlled to track the set value while satisfying the constraint \eqref{state_constraint},  $x_2(k+1|t)$ can be automatically bounded, i.e., $x_2(k+1|t)$ $\in$ [$\alpha_r^{\text{min}}$, $\alpha_r^{\text{max}}$], while tracking an appropriate desired value.} 

By solving the formulated MPC problem, we obtain the following locally optimized control sequence at time $t$:
\begin{equation}\label{control_input_sequence}
    U_N(t)=\left \{u(0|t),... ,u(N-1|t)  \right \}.
\end{equation}
\indent All elements in the obtained control sequence \eqref{control_input_sequence} are constrained between $u^{\text{min}}$ and $u^{\text{max}}$, and only the first element is applied to the vehicle in the MPC framework, i.e., $M_z(t)=u(0|t)$, and all of these processes from \eqref{MPC_problem} to \eqref{control_input_sequence} are repeated at every control cycle to update the new state information.

\subsection{Control Allocation}
The obtained yaw moment from MPC framework $M_z(t)$ can be generated by appropriately allocating the motor torques to the rear left and rear right wheels. The maximum yaw moment that can be created using only a single motor's maximum torque\footnote{We assume the symmetrical motor torque for positive and negative torque areas, i.e., $T_{m,j}^{\text{max}}(v_x)=-T_{m,j}^{\text{min}}(v_x)$, and only the positive torque is exploited when computing the maximum yaw moment for simplicity. Therefore, $M_z^{\text{max}}$ is always positive.} is as follows:
\begin{equation}
    M_z^{\text{max}}=\frac{l_w}{2}\cdot\frac{T_{m,j}^{\text{max}}(v_x)}{r_w}.
\end{equation}
where $j\in\{l,r\}$.
\begin{algorithm}[!t]
\caption{Control Allocation}\label{daisy_chain}
\begin{algorithmic}[1]
\Function{MotorTorque}{$M_{z}(t),\: T_{m,j}^{\text{max}}(v_x)$} $, j\in \left \{l,r \right \}$
    \If{$|M_{z}(t)| \leq M_z^{\text{max}}$}
        \If{$M_z(t)\geq0$}
        \State $T_{m,r} = 2\frac{r_w M_z(t)}{l_w}$ and $T_{m,l} = 0$
        \Else{}
        \State $T_{m,r} = 0$ and $T_{m,l} = -2\frac{r_w M_z(t)}{l_w}$  
    \EndIf
    \Else{}
        \If{$M_z\geq0$}
        \State $T_{m,r} = T_{m,r}^{\text{max}}$ and $T_{m,l} = -2\frac{r_w (M_z(t)-M_z^{\text{max}})}{l_w}$
        \Else{}
        \State $T_{m,r} = -2\frac{r_w (-M_z(t)-M_z^{\text{max}})}{l_w}$ and $T_{m,l} = T_{m,l}^{\text{max}}$
        \EndIf
    \EndIf
\EndFunction
\end{algorithmic}
\end{algorithm}

\indent Since the proposed method aims to improve the vehicle cornering performance, negative torque that decreases the vehicle agility should be minimized. With this in mind, we exploited the daisy chain allocation method \cite{oppenheimer2006control} in this study. The overall strategy is summarized in Algorithm \ref{daisy_chain}. \\
\indent As shown in Fig. \ref{motor_map}, the static motor characteristics curve is utilized to obtain the maximum motor torque according to the vehicle speed. Given a required moment $M_z(t)$ by the MPC framework, only positive motor torque is applied as much as possible to only a single wheel. However, negative torque is also taken into account on the other side wheel, if the required yaw moment exceeds the maximum yaw moment that can be generated by only positive torque, $|M_z(t)|>M_z^{\text{max}}$. For example, if the positive yaw moment is required, the positive motor torque of the right wheel is generated as much as possible, but the negative torque of the left wheel is produced only if required, as explained in Algorithm. \ref{daisy_chain}. \textcolor{black}{In this paper, we assume that the required $M_z(t)$ can be covered by given maximum torques, i.e., $T_{m,l}^{\text{max}}$ and $T_{m,r}^{\text{max}}$ in Fig.~\ref{motor_map}.}

\section{Simulation Results}
In this section, we verify the effectiveness of the proposed approach by simulation studies with the B-class sport vehicle model stored in CarSim, a commercial vehicle dynamics solver package. The model parameter values are summarized in Table \ref{model_parameters}, and the simulations are conducted in the following various test scenarios.
\begin{table}[!h]
\caption{Model Parameter Values.} \label{model_parameters}
\centering
\begin{tabular}{ccc}
\hline \hline
Symbol & Description                                                                                       & Value {[}Unit{]}   \\ \hline
$m$    & vehicle total \textcolor{black}{mass}                                                              & 1140 {[}kg{]}      \\
$l_f$  & \begin{tabular}[c]{@{}c@{}}distance from the center for gravity \\ to the front axle\end{tabular} & 1.165 {[}m{]}      \\
$l_r$  & \begin{tabular}[c]{@{}c@{}}distance from the center for gravity \\ to the rear axle\end{tabular}  & 1.165 {[}m{]}      \\
$C_f$  & front tire cornering stiffness                                                                    & 150000 {[}N/rad{]} \\
$C_r$  & rear tire cornering stiffness                                                                     & 170000 {[}N/rad{]} \\
$r_w$  & tire radius                                                                                       & 0.333 {[}m{]}      \\
$l_w$  & track width                                                                                       & 1.481 {[}m{]}      \\
$g$    & gravitational constant                                                                             & 9.81 {[}m/$\text{s}^2${]}       \\
$T_s$  & sampling time                                                                                     & 5 {[}msec{]}    \\
$N$    & prediction horizon                                                                                & 5 {[}-{]}          \\ \hline                                                                 \hline
\end{tabular}
\end{table}
\subsection{Control performances on different road surfaces}
One of the major advantages of the proposed method is its robustness to varying road surface conditions. To verify this robustness, two test scenarios are designed on different road surfaces. Except for road friction, the other test conditions including vehicle maneuvering are identical for both scenarios.
\begin{figure}[!t]
\centering
\includegraphics[width=3.5in]{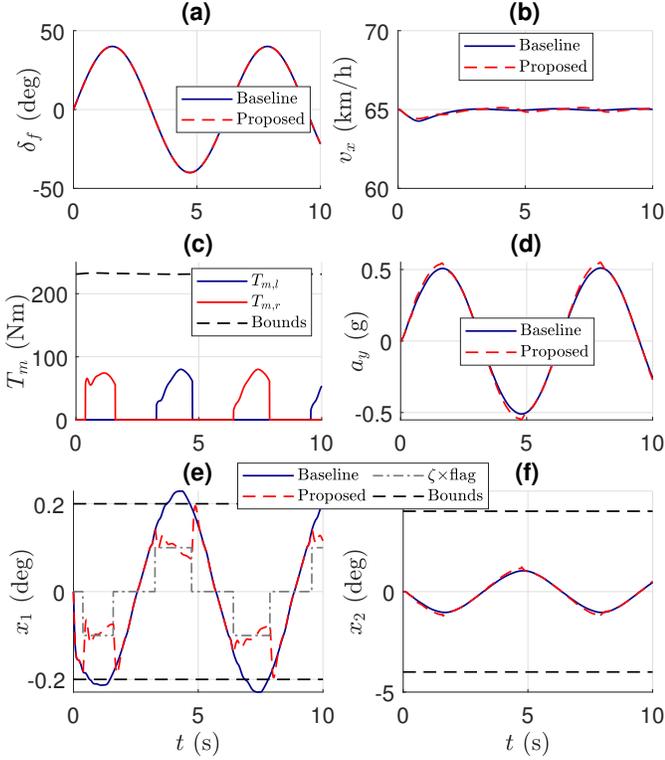}
\caption{\textcolor{black}{Simulation results with mild driver inputs at high-mu road friction.}}
\label{simul1}
\end{figure}
The plots in Fig. \ref{simul1} describe the simulation results at the high-mu road surface with mild steering input, as in Fig. \ref{simul1}(a). The vehicle speed is controlled by an internal speed controller in CarSim to stay at around 65km/h, as shown in Fig. \ref{simul1}(b). 
It should be noted that the appropriate amount of control input is dependent on driver inputs, i.e., steering wheel angle and vehicle speed. Thus, the designed controller does not simply apply excessive torques but rather increases the dynamic performance of the vehicle according to the given driver inputs. In this scenario, the mild steering input and relative low vehicle speed are executed by the driver, so only a small amount of motor torque input is appropriate. Consequently, the resulting torques also satisfy the control constraint, as depicted in Fig. \ref{simul1}(c). 

The control performance is evaluated by the magnitude of lateral acceleration, as shown in Fig. \ref{simul1}(d). We see the increased lateral acceleration for the same steering in \ref{simul1}(a) compared to the uncontrolled case (Baseline). As mentioned, the reference trajectory for the state variable and road friction is not explicitly assigned and given in this paper, so the amount of increased lateral acceleration as shown in Fig. \ref{simul1}(d) is appropriately determined by the driver's inputs. That is, the designed controller can increase the lateral acceleration up to the appropriate magnitude regardless of road type.

In addition, we confirm that the state variable $x_1$, i.e., $\alpha_f -\alpha_r$, is enforced to track the desired gap $\zeta$ when the control flag is turned on while satisfying bound constraints as described in Fig. \ref{simul1}(e). However, the gap between the front and rear wheels is not reduced for the baseline case, and the state variable profile violates the specified constraints. As illustrated in Fig. \ref{simul1}(f), the state variable $x_2$, i.e., $\alpha_r$, is slightly increased due to the small amount of applied torques within acceptable bounds.
\begin{figure}[!t]
\centering
\includegraphics[width=3.5in]{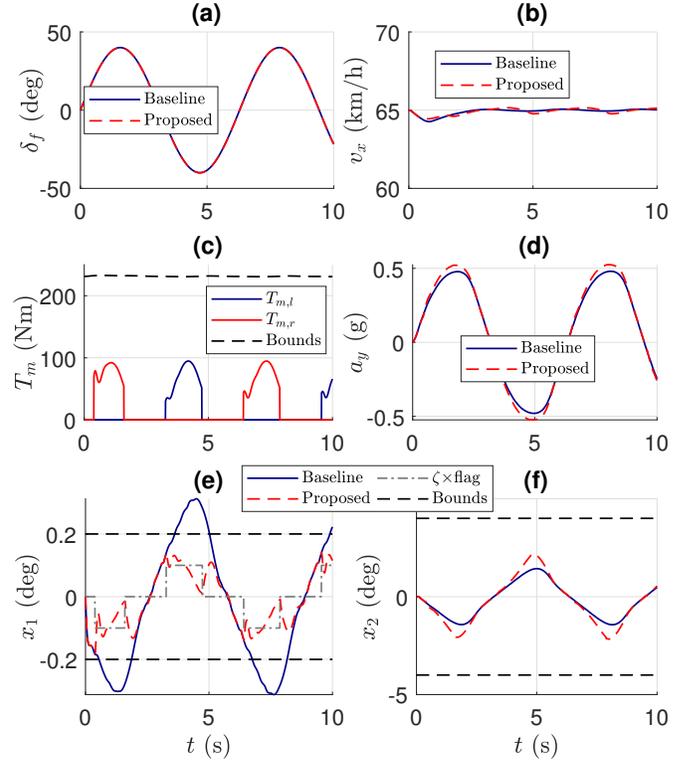}
\caption{\textcolor{black}{Simulation results with mild driver inputs at medium-mu road friction.}}
\label{simul2}
\end{figure}

The robustness to road surface change of a controller with the same performance criteria and the same constraints as in the case of Fig. \ref{simul1} is verified in Fig. \ref{simul2}. In this scenario, the vehicle is maneuvered at the medium-mu road friction. We impose the same driver inputs compared as those for Fig. \ref{simul1}, as shown in Fig. \ref{simul2}. Compared with the previous case, more vehicle slippage is observed due to the reduced road friction, as illustrated in Figs. \ref{simul2}(e) and \ref{simul2}(f). By enforcing the state variables to satisfy the constraints in the MPC framework, we verify that the vehicle is controlled to turn dynamically by applying appropriate motor torques to individual wheels. 

Table \ref{control_performance} shows the overall control performances in terms of the maximum lateral acceleration, which is increased around 7$\%\sim9\%$ for both cases. Note that our approach produces a larger $a_y$ at the slippage than the $a_y$ at the high-mu for the baseline. Nevertheless, we cannot conclude that the proposed MPC is the best controller to improve cornering performance. It is very apparent that additional yaw moment generated by the MPC increases the lateral acceleration to some extent. Therefore, comparing it with the uncontrolled results are unfair to confirm the benefit of the proposed controller. Therefore, we compare the MPC with the conventional controller in next test scenario. This sub-section, however, verifies that the proposed MPC can handle several specified constraints, which can prevent vehicle instability. In addition, we confirm that the proposed MPC is applicable regardless of road surface condition.
\begin{table}[!h]
\centering
\caption{Control performances in terms of Maximum Lateral Acceleration.} \label{control_performance}
\begin{tabular}{cccc}
\hline \hline
 & \begin{tabular}[c]{@{}c@{}}Max $|a_y|$ {[}g{]}\\ (baseline)\end{tabular} & \begin{tabular}[c]{@{}c@{}}Max $|a_y|$ {[}g{]}\\ (proposed)\end{tabular} & Improvement {[}\%{]} \\ \hline Fig. \ref{simul1}
 & 0.5069                                                                      & 0.5456                                                                   & 7.09                     \\  Fig. \ref{simul2}
 & 0.4758                                                                      & 0.5206                                                                   & 8.61                      \\ \hline \hline
\end{tabular}
\end{table}

\subsection{Comparison between proposed and conventional controllers}
Compared to conventional feedback error-based controllers, the most distinguishing feature of MPC is that it can handle the constraints by predicting the model deployment during a specified prediction horizon. To show this strength, we compare proposed MPC with the conventional controller.

For the purpose of comparison, the following conventional controller designed in a such a way as to react immediately to sideslip difference feedback error is introduced.

The control law should satisfy the following condition for the asymptotic stability: 
\begin{equation}\label{smc_objective}
    \dot{s}=-\lambda s.
\end{equation}
where $s$ is the sliding surface that is defined by \textcolor{black}{$s=\text{sign}\left (\alpha_f - \alpha_r\right)\left (\alpha_f - \alpha_r\right ) -\zeta$} (sideslip difference feedback error), and $\lambda$ is a non-negative control gain.

\textcolor{black}{Substituting the developed model in \eqref{state_space} into \eqref{smc_objective} leads to, }
\begin{equation}\label{smc_derivation}
  \textcolor{black}{  \text{sign}\left(x_1\right)\dot{x}_1=-\lambda s,}
\end{equation}
\textcolor{black}{and $x_1=\alpha_f-\alpha_r$, that is defined by \eqref{eq:alpha_diff}.}

\textcolor{black}{Now, the following control law can be derived by substituting \eqref{eq:alpha_diff} and \eqref{eq:bicyle_model2} into \eqref{smc_derivation}:}
\begin{align}
    &\textcolor{black}{\text{sign}\left(x_1\right)\left\{ \frac{L}{v_{x}}\dot{r}-\dot{\delta}_{f}\right\}=-\lambda s,\nonumber}\\
    &\textcolor{black}{\text{sign}\left(x_1\right)\left\{ \frac{L}{v_{x}}\left(\frac{l_{f}F_{yf}\text{cos}\delta_f-l_{r}F_{yr}+M_{z}}{I_{z}}\right)-\dot{\delta}_{f}\right\}=-\lambda s,\nonumber}
\end{align}

\textcolor{black}{To satisfy stability condition in \eqref{smc_objective}, the control law $u$ is derived as,
\begin{align}\label{eq:conv_control}
    M_z&=-l_fF_{yf}\text{cos}\delta_f+l_rF_{yr}+
        \frac{I_zv_x}{L}\dot{\delta}_f-\frac{I_zv_x}{L\cdot \text{sign}(x_1)}\lambda \cdot s, \nonumber\\
        &\approx-l_fF_{yf}+l_rF_{yr}+
        \frac{I_zv_x}{L}\dot{\delta}_f-\text{sign}(x_1)\frac{I_zv_x}{L}\lambda \cdot s
\end{align}
Here, we assume the small steered wheel angle.}

We denote this controller the conventional control method. 
To compare the conventional controller and the proposed MPC in a fair manner, we use the same values of the desired set value and control allocation method. Moreover, the conventional controller's gain is tuned to their full potential for optimum control performances. However, it is obvious that the designed control in \eqref{eq:conv_control} lacks the ability to predict the model deployment and handle the constraints.
\begin{figure}[!t]
\centering
\includegraphics[width=3.5in]{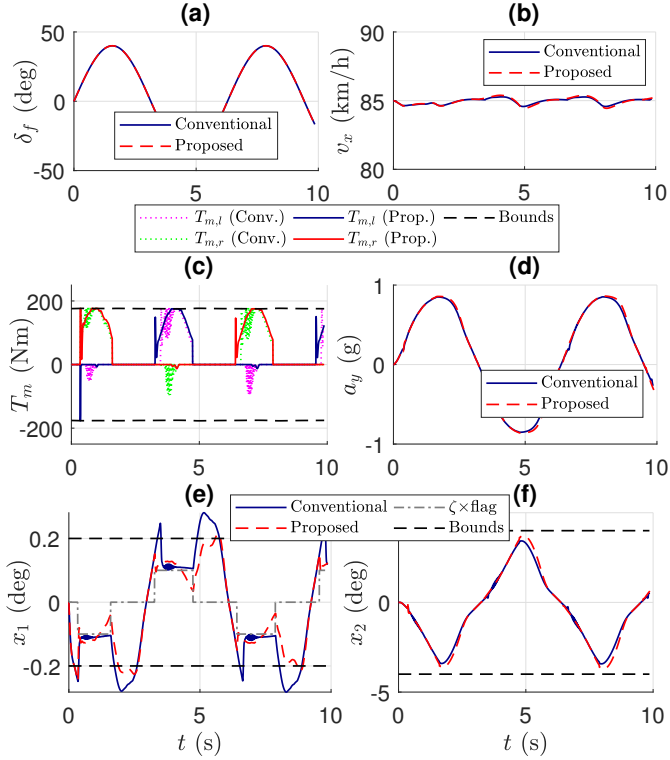}
\caption{\textcolor{black}{Comparison of simulation results between the developed MPC and the conventional controller at high-mu road friction.}}
\label{simul3}
\end{figure}
Figure \ref{simul3} compares the simulation results for the conventional and proposed methods. To verify the ability to handle constraints, the vehicle is more aggressively maneuvered than in the test scenarios in the previous sub-section but the same steering input is applied, as illustrated Figs. \ref{simul3}(a) and \ref{simul3}(b). Therefore, control input reaches its allowable bound, as described in Fig. \ref{simul3}(c). 
Since the developed MPC framework can predict the model deployment for a specified time horizon, the controller immediately decreases its input magnitude to prevent vehicle instability, as shown in Fig. \ref{simul3}(c). However, the conventional method shows severe oscillation when it reaches the control limit due to the accumulated feedback error, which eventually causes the vehicle instability. By elaborating the control gain tuning, we can avoid the vehicle becoming completely unstable, but it may easily lose stability when unexpected uncertainties come from the environment. Fig. \ref{simul3}(d) shows that the similar lateral acceleration trajectories for both controllers. However, since the conventional controller is designed to react to feedback error immediately, the vehicle is actually at the border between stable and unstable areas when the actuator reaches to its limit. In contrast, the proposed MPC enforces the state variable within allowable limits in advance by predicting model deployment and gives the locally optimal control for the given environments (driver's input and road friction), see Figs. \ref{simul3}(c) - \ref{simul3}(f).

\textcolor{black}{From this simulation results, it cannot be concluded that the MPC always outperforms other controllers in terms of cornering performance, see Fig.~\ref{simul3}(d). However, although conventional control can show better or similar control performance by tuning the gains in a such a way as to minimize the feedback error as much as possible, excessive control without considering constraints may cause control saturation, which eventually leads to vehicle instability. In contrast, multiple aspects such as state and control constraints can be handled by the MPC simultaneously, so the vehicle can be maneuvered with a certain amount of safety margin. In that sense, the proposed method with MPC framework is a better choice than a conventional controller, although control performance is not always better.}
\subsection{Path following with the driver model}
\begin{figure}[!t]
\centering
\includegraphics[width=3.5in]{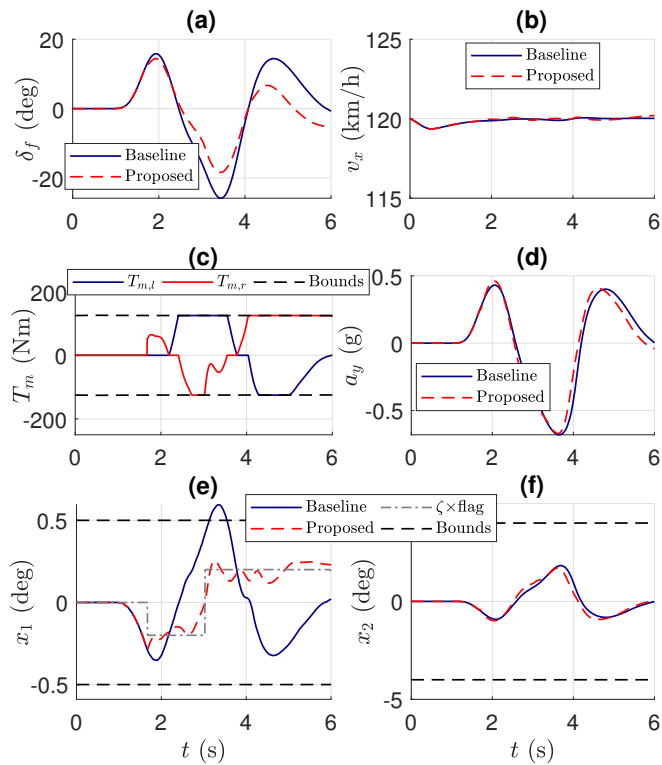}
\caption{\textcolor{black}{Simulation results with the driver model in a double lane change scenario at high-mu road friction.}}
\label{simul4}
\end{figure}
In the previous test scenarios, to guarantee the reproduciblity of test conditions and environments, the same steering inputs were applied for all cases, whether the vehicle followed the desired path or not. However, in reality, the driver tries to maneuver the vehicle according to changes in the road profile. In other words, keeping in mind that the vehicle should follow the desired path, the driver applies the appropriate steering, accelerator, and barke pedal inputs to the vehicle in real-time. Therefore, the designed controller in this paper needs to adaptively provide corrective yaw moment according to the driver's intention and maneuvers.

To verify this ability, we employ the driver model stored in CarSim. With this model, regardless of control inputs, the driver always tries to follow the defined trajectory (double lane change maneuvering) as much as possible unless it is sometimes physically difficult to track the path.

As shown in Fig. \ref{simul4}(a), unlike the previous scenarios, the applied steering wheel angle by the driver model for the baseline and proposed cases are different. We can see that a smaller steering input is applied to the proposed case. Nevertheless, the vehicle exhibits a larger or similar lateral acceleration assisted by the corrective yaw moment, as described in Figs. \ref{simul4}(c) and \ref{simul4}(d). By enforcing the state variables within the specified bounds, the vehicle is also prevented from losing its stability, as shown in Figs. \ref{simul4}(e) and \ref{simul4}(f). In this scenario, compared to previous cases, more relaxed constraints are specified due to the severe maneuvering, and these variations of constraints according to the driver maneuvers are reasonable and possible in real-world applications. \textcolor{black}{Through this verification, we confirm that the developed controller fulfills its role faithfully in collaboration with driver who maneuvers the vehicle adaptively depending on the driving conditions.}

\section{Conclusions}
In this paper, we present a new control strategy to improve the vehicle cornering performance in a model predictive control framework. The main contributions of this paper that distinguish it from other studies are two fold. (i) By exploiting vehicle natural handling characteristics, the optimal corrective yaw moment is determined in such a way as to utilize the road friction as much as possible without defining the desired yaw rate. Furthermore, the proposed method is effective for any road friction.  (ii) A new mathematical model that predicts the behavior of the wheel sideslip difference between the front and rear wheels is established. Using this model, the controller is designed to keep the constant gap of the wheel sideslip difference to improve the cornering performance. Moreover, the state and control are enforced to satisfy the constraints, which enables the vehicle to be appropriately controlled without losing its stability. We confirmed these advantages through the various simulations, and we believe that the proposed method creates a new possibility to reduce the complexity of the existing heuristic control methods.


%



\ifCLASSOPTIONcaptionsoff
  \newpage
\fi



%

\bibliographystyle{IEEEtran}
\bibliography{ref}

%

\begin{IEEEbiography}
    [{\includegraphics[width=1in,height=1.25in,clip,keepaspectratio]{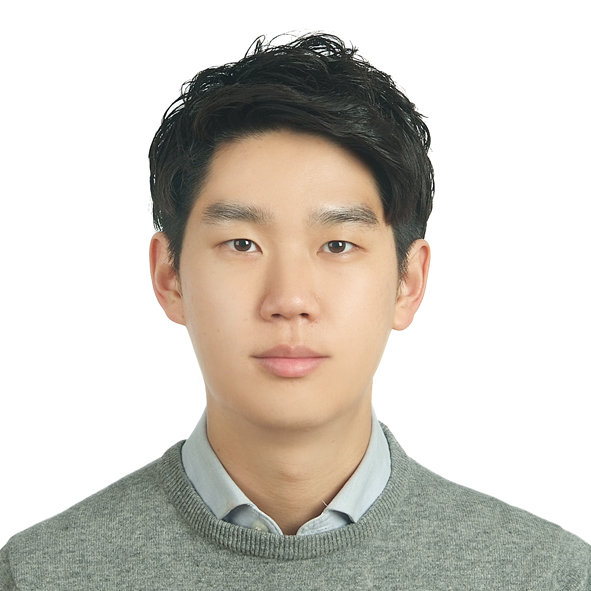}}]{Kyoungseok Han}
received the B.S. degree in civil engineering with a minor in mechanical engineering from Hanyang University, Seoul, Korea in 2013 and the M.S. and Ph.D. degrees in mechanical engineering from the Korea Advanced Institute of Science and Technology (KAIST), Daejeon, Korea in 2015 and 2018. He is currently a postdoctoral research fellow at the University of Michigan, Ann Arbor, MI, USA. He mainly wrote this paper while studying as a Ph. D. candidate at KAIST, and concluded the draft during his postdoctoral research fellow period. His current research interests include vehicle dynamics and control, electrified autonomous vehicles, reinforcement learning, and optimization.
\end{IEEEbiography}
\vskip 0pt plus -1fil
\begin{IEEEbiography}
    [{\includegraphics[width=1in,height=1.25in,clip,keepaspectratio]{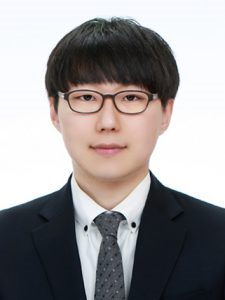}}]{Giseo Park}
received the B.S. degree in mechanical engineering from Hanyang University, Seoul, Korea, in 2014 and the M.S. degree in mechanical engineering from Korea Advanced Institute of Science and Technology (KAIST), in 2016. He is currently a Ph.D. candidate of mechanical engineering department at KAIST. His research
interests include vehicle dynamics, control theory, active safety systems and autonomous driving system.
\end{IEEEbiography}
\vskip 0pt plus -1fil
\begin{IEEEbiography}
    [{\includegraphics[width=1in,height=1.25in,clip,keepaspectratio]{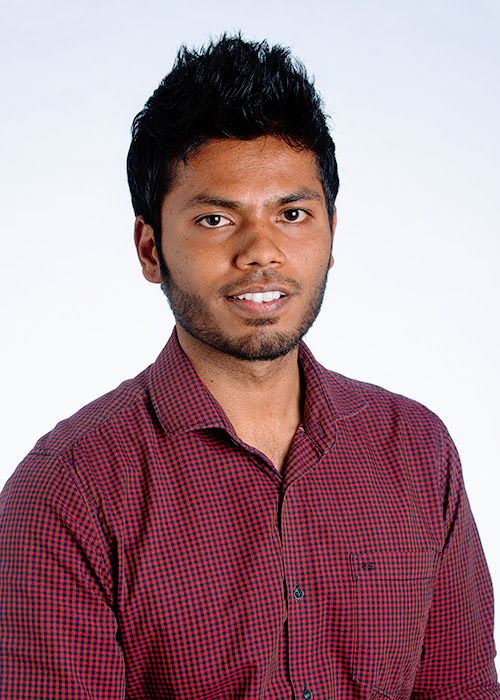}}]{Gokul S. SanKar} pursued his Ph.D. at the University of Melbourne, Parkville, VIC, Australia, and received his M.S. from Indian Institute of Technology (Madras), India, in 2013 and Bachelor’s degree in Electronics $\&$ Instrumentation Engineering, from Anna University, India, in 2010.
    
He is currently a Research Fellow with the University of Michigan, Ann Arbor, MI, USA. His research interests include model predictive control with applications 
to automotive systems, autonomous vehicles, and model-based robust control algorithms.

He was a member of exclusive Melbourne-India postgraduate program (MIPP) cohort at the University of Melbourne during his doctoral studies.
\end{IEEEbiography}
\vskip 0pt plus -1fil
\begin{IEEEbiography}
    [{\includegraphics[width=1in,height=1.25in,clip,keepaspectratio]{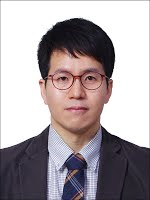}}]{Kanghyun Nam}
received the B.S. degree in mechanical engineering from Kyungpook National University, Daegu, South Korea, in 2007, the M.S. degree in mechanical engineering from Korea Advanced Institute of Science and
Technology, Daejeon, South Korea, in 2009, and the Ph.D. degree in electrical
engineering from The University of Tokyo, Tokyo, Japan, in 2012. From 2012 to 2015, he was a Senior Engineer with Samsung Electronics Co., Ltd., Gyeonggido,
South Korea. Since 2015, he has been an Assistant Professor in the School of Mechanical Engineering, Yeungnam University, Gyeongsan, South Korea. His research interests include motion control, and vehicle control.
\end{IEEEbiography}
\vskip 0pt plus -1fil
\begin{IEEEbiography}
    [{\includegraphics[width=1in,height=1.25in,clip,keepaspectratio]{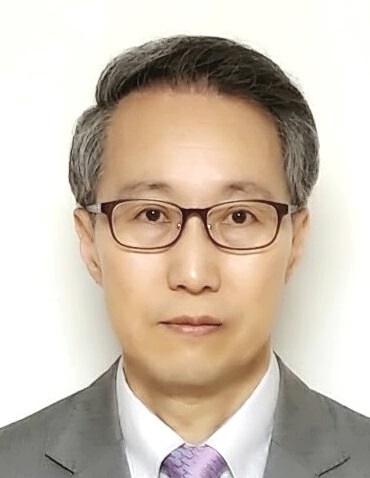}}]{Seibum B. Choi(M'09)}
received the B.S. degree in mechanical engineering from Seoul National University, Seoul, Korea, in 1985, the M.S. degree in mechanical engineering from the Korea Advanced Institute of Science and Technology (KAIST), Daejeon, Korea in 1987, and the Ph.D. degree in control from the University of California, Berkeley, CA, USA, in 1993. From 1993 to 1997, he was involved in the development of automated vehicle control systems at the Institute of Transportation Studies, University of California. During 2006, he was with TRW, Warren, MI, USA, where he was involved in the development of advanced vehicle control systems. Since 2006, he has been with the faculty of the Mechanical Engineering Department, KAIST. His current research interests include fuel-saving technology, vehicle dynamics and control, and active safety systems. 
\end{IEEEbiography}






\end{document}